\documentclass[preprint,showpacs,preprintnumbers,amsmath,amssymb]{revtex4}

\usepackage{graphicx}
\usepackage{dcolumn}
\usepackage{bm}
\usepackage{amssymb}
\usepackage{graphicx}
\usepackage{amsmath}
\usepackage{xspace}

\begin{document}

\title{Raman study of Twin Free YBa$_2$Cu$_3$O$_{6.5}$ (Ortho-II) Single Crystals }
\author{M. N. Iliev,  V.~G.~Hadjiev}
\affiliation{Texas Center for Superconductivity, University of Houston, Texas 77204-5002, USA}
\author{S. Jandl, D.~Le Boeuf}
\affiliation{Regroupement Qu\'ebecois sur les Mat\'eriaux de
Pointe, D\'epartement de Physique, Universit\'e de Sherbrooke,
Sherbrooke, Canada J1K 2R1}
\author{V. N. Popov}
\affiliation{Faculty of Physics, University of Sofia, 1164 Sofia, Bulgaria}
\author{D. Bonn, R. Liang, W. N. Hardy}
\affiliation{Department of Physics and Astronomy, University of
British Colombia,  Vancouver,BC, Canada V6T 1Z1}
\date{\today}

\begin{abstract}
The polarized Raman scattering spectra from freshly cleaved $ab$,
$ac$, and $bc$ surfaces of high quality twin free
YBa$_2$Cu$_3$O$_{6.5}$ (Ortho-II) single crystals ($T_c$=57.5~K
and $\Delta T = 0.6$~K) were studied between 80 and 300~K. All
eleven $A_g$ Raman modes expected for the Ortho-II structure as
well some modes of $B_{2g}$ and $B_{3g}$ symmetry were identified
in close comparison with predictions of lattice dynamical
calculations. The electronic scattering from the $ab$ planes is
strongly anisotropic and decreases between 200 and 100~K within
the temperature range of previously reported pseudogap opening.
The coupling of phonons to Raman active electronic excitations
manifested by asymmetric (Fano) profiles of several modes also
decreases in the same range. Among the new findings that
distinguish the Raman scattering of Ortho-II from that of Ortho-I
phase is the unusual relationship ($\alpha_{xx} \approx
-\alpha_{yy}$) between the elements of the Raman tensor of the
apex oxygen $A_g$ mode.
\end{abstract}

\pacs{78.30.-j, 74.25.Kc, 74.72.Bk}

\maketitle

\section{Introduction}
The properties of underdoped YBa$_2$Cu$_3$O$_x$ ($6 < x < 7$) have
been studied intensively within the efforts to unravel the
mechanism of high temperature superconductivity. The Ortho~II
phase, corresponding to oxygen content $x = 6.5$ and characterized
by alternating full and empty chains, has attracted particular
attention as it is both underdoped and free of disorder. Although
the existence of this phase has been documented experimentally in
considerable number of reports, it has also been established that
as a rule the Ortho~II domains coexists with domains of different
oxygen ordering or/and strong disorder even in the case $x\simeq
6.5$.
An improved procedure developed by Liang at al.\cite{liang2000}
has made possible preparation of twin free, highly ordered
Ortho~II single crystals. Such crystals have recently been used in
several studies of specific properties of the Ortho~II phase by
means of neutron scattering\cite{stock2002,stock2004}, x-ray
diffraction\cite{zimmermann2003}, time-resolved
spectroscopy\cite{gedik2004,gedik2004a}, infrared
spectroscopy\cite{hwang2006}, nuclear magnetic
resonance\cite{yamani2006}, microwave
spectroscopy\cite{harris2006}, and resistance measurements in high
magnetic fields.\cite{leyraud2007}.

There have been several attempts to identify the Raman modes of the
Ortho-II phase by measuring the spectra of oxygen deficient
YBa$_2$Cu$_3$O$_x$ ($x\approx 6.5$) single
crystals.\cite{iliev1993,misochko1994,iliev1996}  Compared to the
well known Raman spectra of Ortho-I ($x=7$) and T ($x=6$) phases, in
the case of Ortho-II one expects shift of the corresponding Raman
modes and activation of additional modes due to the doubling of the
unit cell. A significant number of additional modes have been
observed, but their identification has met definite difficulties due
to ambiguities in the local structure of YBa$_2$Cu$_3$O$_x$ even in
the case $x=6.5$. Indeed, in the idealized Ortho-I, Ortho-II and T
structures the Cu1 and O1 atoms in the basal Cu-O planes are at
centrosymmetrical sites and their vibrations are not Raman active.
The oxygen arrangement in the basal planes of a real
YBa$_2$Cu$_3$O$_x$ material, however, is characterized by chain
fragments instead of infinite chains and part of oxygen atoms  are
outside the chains in otherwise vacant O5 sites. This creates a
number of local non-centrosymmetrical surroundings for Cu and O
atom, in particular those at the end of chain fragments, and
activates their vibrations in the Raman
spectrum.\cite{iliev1996,falques1997} Some of these defects can also
be produced by local laser annealing\cite{iliev1997,iliev1997a} or
photoactivation.\cite{osada2005} Another issue, as a rule neglected
but of particular  importance in the case of Ortho-II phase, is the
possible lack of correspondence between the oxygen content and
arrangement in the volume of the crystal and its surface layer(s),
where the Raman scattering occurs. The cause for such discrepancy is
the in- and out-diffusion of oxygen, which depends strongly on the
type of the surface ($ab$, $bc$ or $ac$), starting oxygen content,
ambient oxygen pressure, temperature and exposure
time.\cite{conder2000}

In this paper we present results of a temperature-dependent Raman
study on freshly-cleaved $ab$, $ac$, and $bc$ surfaces of a twin
free YBa$_2$Cu$_3$O$_{6.5}$ (YBCO6.5) single crystal with high
degree of Ortho-II type ordering. This allowed us with great
certainty to identify the proper Raman modes of the Ortho-II phase
and assign them to definite atomic motions via close comparison with
predictions of lattice dynamical calculations, as well as to measure
the symmetry and strength of electronic scattering.

\section{Samples and Experimental}
We used a high quality mechanically de-twinned Ortho-II YBCO6.5
single crystal with $T_c$=57.5~K and $\Delta T =0.6~K$, grown by a
flux method in BaZrO3 crucible.\cite{liang2000,leyraud2007}
Immediately before mounting the sample on the cold finger of a
Microstat$\circledR$He (Oxford Instruments) optical cryostat, a
small area of the surface ($ab$, $ac$, or $bc$) to be used for Raman
measurements was cleaved out to ensure that the scattering volume
has an Ortho-II oxygen arrangement. The Raman spectra were measured
under microscope ($\times 50$ magnification) using a triple T64000
(Horiba Jobin Yvon) spectrometer. In most experiments we used 633~nm
excitation with less than 1~mW incident laser power focused at a
spot of 2-3~$\mu$m diameter. Comparative measurements with 515, 488,
and 458~nm excitations were also done. All spectra were corrected
for the Bose factor. For description of the scattering
configurations we use the Porto's notation $a(BC)d$, where the first
and fourth letters denote, respectively, the directions of incident
and scattered light in a Cartesian $xyz$ system with axes along the
crystallographic directions. The polarization of the incident and
scattered light is given by the second and third letters,
respectively.

\section{Results and Discussion}

\subsection{Raman phonons in the Ortho-II phase}
In Figure 1 are compared the Raman spectra obtained with 633~nm
excitation at room temperature from freshly cleaved and aged areas
on the $ab$ and $bc$ surfaces of the same Ortho-II crystal. The
spectral shapes and Raman line frequencies from aged surfaces are
consistent with those from earlier reports on the $XX/YY$ and $ZZ$
spectra of twinned YBCO6.5\cite{misochko1994} and YBCO$x$ ($x\approx
6.5$ samples.\cite{palles1996,hong2007} The corresponding spectra from
freshly cleaved and aged surfaces, however, exhibit definite
differences in the positions and appearance of some of the peaks.
\begin{figure}[htbp]
\includegraphics[width=8cm]{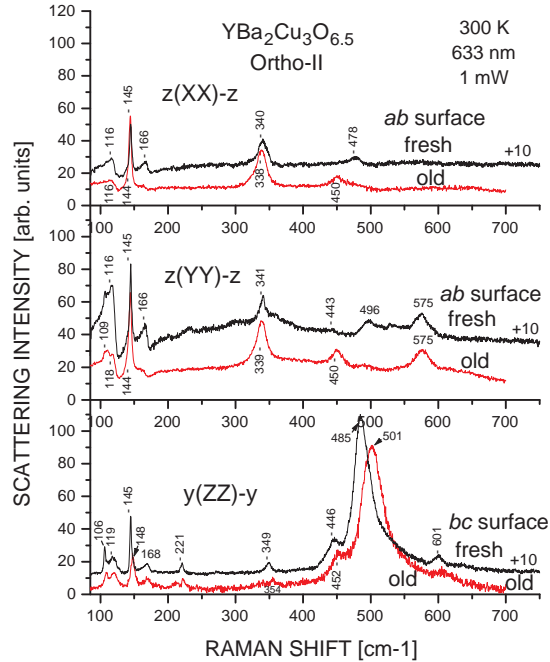}
\caption{(Color online) Raman spectra of YBa$_2$Cu$_3$O$_{6.5}$
(Ortho-II) obtained from freshly cleaved and aged $ab$ and $bc$
surfaces.}
\end{figure}

Figure 2 shows the Raman spectra from freshly cleaved surfaces at
90~K in all available exact scattering configurations. From symmetry
considerations one expects for the Ortho-II structure an increased
number of Raman active modes, $11A_g + 4B_{1g} + 11B_{2g} +
8B_{3g}$, compared to the $5A_g + 5B_{2g} + 5B_{3g}$ modes of the
Ortho-I structure. The atomic displacements of the eleven $A_g$
modes, as predicted by lattice dynamical calculations,
(LDC)\cite{iliev1996} are shown in Figure 3. For most modes, the
experimentally observed frequencies (at 90~K) are in good agreement
with those predicted by LDC, shown in parenthesis. In addition to
the five $A_g$ modes at 126, 147, 342, 447, and 487~cm$^{-1}$
corresponding to displacements along the $c$-axis of Ba, Cu2, O2-O3,
O2+O3, O4 in the Ortho-I structure, there are six more $A_g$ modes
in the extended Ortho-II cell. Four of them  can be positively
identified at 107~cm$^{-1}$ [Cu2($z$)-Cu2'($-z$) out-of-phase],
171~cm$^{-1}$ [Y($x$)], 352~cm$^{-1}$ [O4($z$)-O4'($-z$)
out-of-phase], and 579~cm$^{-1}$ [O2($x$)]. The mixed Ba/Cu mode,
predicted at 146~cm$^{-1}$, may be very weak and not observable. The
weak Raman line at 381~cm$^{-1}$ can be tentatively assigned to the
mode involving mainly O4' displacements along $c$, with a predicted
frequency of 414~cm$^{-1}$. In the $ZZ$ spectra one observes
additional lines of $A_g$ character at 224 and 601~cm$^{-1}$. These
positions are close to those of defect modes related to
displacements of Cu1 and O1 at the end of broken
chains,\cite{iliev1996,iliev1997,hong2007} which could preexist or
be created by photoactivation in our Ortho-II sample.

\begin{figure}[htbp]
\includegraphics[width=8cm]{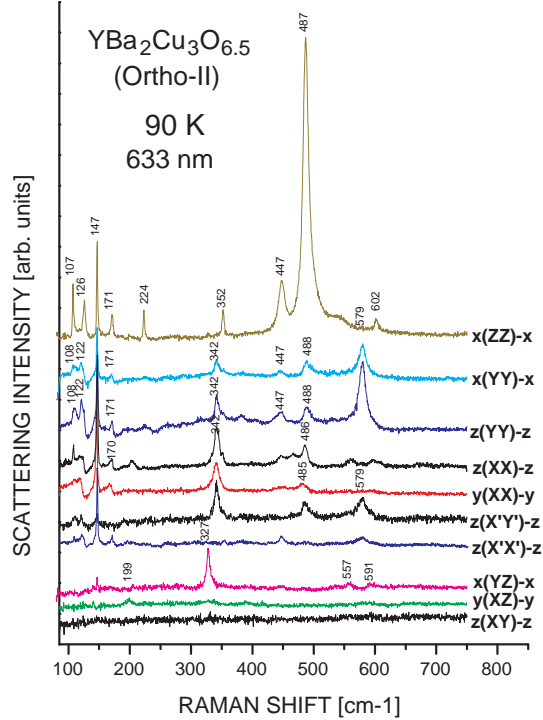}
\caption{(Color online) Raman spectra of YBa$_2$Cu$_3$O$_{6.5}$
(Ortho-II) obtained at 90~K from freshly cleaved $ab$, $ac$ and $bc$
surfaces. The spectra are shifted vertically for clarity.}
\end{figure}

\begin{figure}[htbp]
\includegraphics[width=8cm]{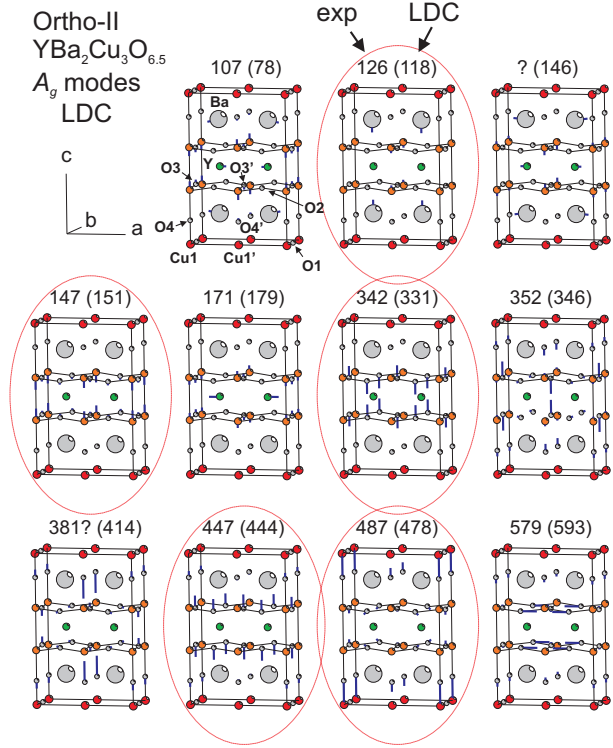}
\caption{(Color online) Main atomic displacements of the $A_g$ modes
of YBa$_2$Cu$_3$O$_{6.5}$ (Ortho-II) as obtained by LDC. The LDC predicted
frequencies (in parenthesis) are compared to experimentally obtained values.}
\end{figure}

In the $XZ$($B_{2g}$) and $YZ$($B_{3g}$) spectra one observes
relatively strong Raman peaks at 199 and 327~cm$^{-1}$,
respectively. The closest $B_{2g}$ (200 and 213~cm$^{-1}$) and
$B_{3g}$ (300~cm$^{-1}$) frequencies predicted by LDC, correspond to
modes involving mainly displacements of O4 and O4' along $a$ and $b$
directions, respectively.

A more careful look at the appearance of the apex oxygen mode near
487~cm$^{-1}$ in the $X'X'$ and $X'Y'$ spectra of Fig. 2 reveals a
significantly different behavior compared to that in the
corresponding spectra of the Ortho-I phase. Indeed, with these
scattering configurations, the intensity of an $A_g$ Raman line is
proportional to  $(\alpha_{xx} + \alpha_{yy})^2$ and $(\alpha_{xx} -
\alpha_{yy})^2$, respectively, where $\alpha_{xx}$ and $\alpha_{yy}$
are diagonal elements of the corresponding Raman tensors. Negligible
intensity in the $X'X'$ spectra may be expected if $\alpha_{xx}
\approx -\alpha_{yy}$, which in the Ortho-I phase is satisfied only
for the out-of-phase O2-O3 mode at $\approx 336$~cm$^{-1}$, but not
for the apex oxygen mode near 500~cm$^{-1}$ (seen in the $X'X'$ but
not in the $X'Y'$ spectrum). Here, for the Ortho-II phase the apex
oxygen mode has an appearance similar to that of the out-of-phase
mode: it is practically not seen in the $X'X'$ spectrum, but well
pronounced in the $X'Y'$ spectrum with an intensity comparable to
that in the $XX$ and $YY$ spectra. This allows us to conclude that
for the apex oxygen mode of the Ortho-II phase, the relation
$\alpha_{xx} \approx -\alpha_{yy}$ is satisfied.

\subsection{Electronic scattering and electron-phonon interaction}

In addition to discrete phonon lines, the Raman spectra of Ortho-II
phase contain structureless background with intensity stronger with
$YY$ than with $XX$, and negligible intensity with $ZZ$ polarization
(Fig.4). Such background scattering, observed in the normal and
superconducting states of high $T_c$ superconductors has been
attributed to electronic scattering and has been intensively studied
both experimentally and theoretically.\cite{strohm1997,
bock1999,opel2000,devereaux2007} It has been shown that for
optimally doped YBCO the electronic scattering is practically
independent of temperature for $T > T_c$. With the opening of a
superconducting gap $\Delta$ at $T < T_c$, there is a redistribution
of electronic scattering intensity from lower to higher energies and
a maximum associated with pair-breaking is formed at $\omega_{max} =
2\Delta$. For underdoped YBCO a slight decrease of electronic
scattering intensity at $\omega < 600$~cm$^{-1}$ has been observed
below a characteristic temperature $T^*$ well above $T_c$ and this
has been considered a manifestation of a pseudogap
opening.\cite{opel2000} There are to our knowledge no reports on the
variation of electronic scattering near $T^*$.

With few exceptions,\cite{krantz1995,strohm1997} in experimental
studies of the electronic scattering in YBCO, twinned samples have
been used and the theoretical models are based on the tetragonal
approximation for the crystal structure.

\begin{figure}[htbp]
\includegraphics[width=8cm]{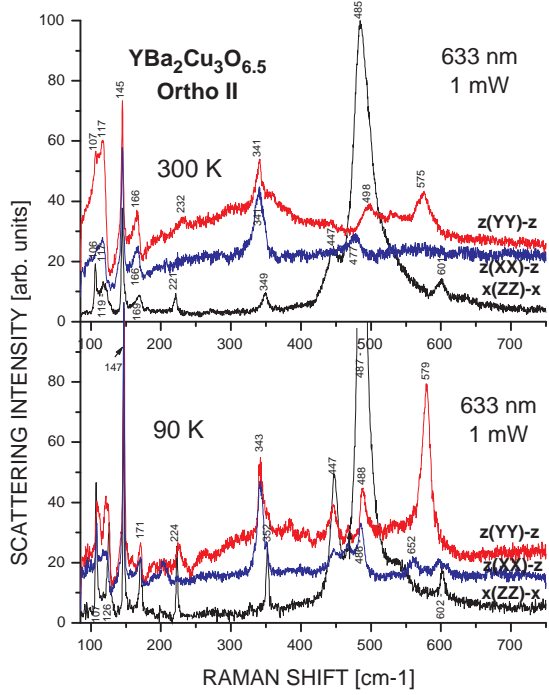}
\caption{(Color online) $z(XX)\bar{z}$, $z(YY)\bar{z}$,  and
$y(ZZ)\bar{y}$ spectra of Ortho-II at 300 and 90~K. Note the
asymmetric Fano shape of the Ba(117~cm$^{-1}$), Y(166~cm$^{-1}$),
and apex oxygen(477 and 498~cm$^{1}$) modes in the $XX$ and $YY$
spectra at 300~K.}
\end{figure}

Figs. 4 and 5 illustrate that the electronic scattering from the
$ab$ surface of the Ortho-II crystal is strongly anisotropic and
temperature dependent below 600~cm$^{-1}$. Its intensity in the
$z(YY)\bar{z}$ spectra is stronger by a factor 2 than in the
$z(XX)\bar{z}$ and much stronger than in the $z(XY)\bar{z}$,
$z(X'Y')\bar{z}$, $x(ZZ)\bar{x}$, and $x(YZ)\bar{x}$ spectra. We
attribute the stronger electronic Raman intensity for $YY$
polarization to the additional scattering channels involving two
CuO$_2$ planar and a CuO chain bands crossing the Fermi surface
along the $\Gamma\,(0,0,0)$ - $Y\,(0,\pi,0)$ direction\cite{bandstr}
in the Brillouin zone .

 \begin{figure}[htbp]
\includegraphics[width=8cm]{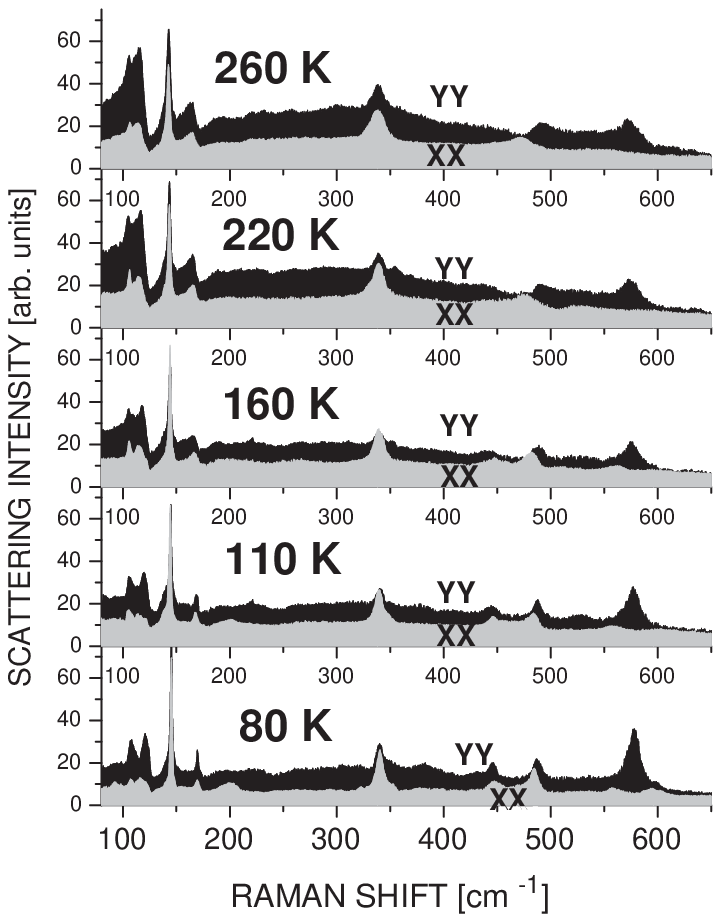}
\caption{(Color online) Variation of the electronic background
in the $z(XX)\bar{z}$ and  $z(YY)\bar{z}$ spectra
of Ortho-II between 260 and 80~K.}
\end{figure}
The coupling of the phonons to the electronic excitations
contributing to the Raman background is manifested in our spectra by
the asymmetric Fano profile of some of the $A_g$ phonon lines, most
clearly pronounced for these
 near 118~cm$^{-1}$(Ba), 168~cm$^{-1}$ (Y) and
486~cm$^{-1}$ (apex oxygen), which interestingly include mainly
motions of atoms not belonging to the Cu-O planes. For a phonon
coupled to an electronic background the Fano profile
\begin{equation}
I(\omega)=I_0\frac{(\epsilon + q)^2}{(1+\epsilon^2)}+B(\omega)
\end{equation}
is generally used to describe the line shape, where $\epsilon =
(\omega - \omega_p)/\Gamma$, $\omega_p$ is the renormalized phonon
frequency that includes all contributions resulting from the
interaction of the phonon with elementary excitations, $\Gamma$ is
the linewidth, $q$ is the asymmetry parameter, and $B(\omega)$ is
the non-interacting with the phonon part of the electronic
excitations continuum. In the case of real phonon and electronic
scattering amplitudes $t_p$ and $t_e$, and a flat scattering
background around the phonon frequency
\begin{equation}
\frac{1}{q} = \frac{t_e}{t_p} \pi \rho V,
\end{equation}
where $\rho(\omega)={\rm const}$ is the density of electronic
excitations coupled to the particular phonon, $V$ is the related
electron-phonon coupling constant, and the intensity of the
interacting with the phonon part of the electronic continuum in
Eq.(1) can be expressed as $I_0=\pi \rho t_e^2$. Under the
reasonable assumption that $t_e$, $t_p$, and $V$ are only weakly
dependent on $T$, the variations with temperature of the absolute
value of $1/q$ and $I_0$ will be governed mainly by $\rho(T)$.
Fig.~6 shows in more detail the variations between 260 and 80~K of
profile of the Raman line near 485~cm$^{-1}$, which corresponds to
the apex oxygen $A_g$ mode. In contrast to the known results for
twin free Ortho-I crystals, the slopes of Fano shaped profiles in
the $XX$ and $YY$ spectra of Ortho-II are of opposite sign, which is
not surprising if we take into account that $t_p^{xx} \propto
\alpha_{xx}$ and $t_p^{yy} \propto \alpha_{yy}$ are of opposite sign
for this mode. It is also evident from Fig. 6 that  the Raman lines
become more symmetric at lower temperature. The Fano fit of the
experimental profiles showed that the $1/q$ factor for the $XX$ and
$YY$ decreases, respectively, from -0.6 and +0.8 at 300~K to -0.2
and +0.1 at 80~K. On the other hand, the electronic background
$B(\omega,T)$ remains nearly constant between 300 and 200~K but with
further cooling decreases significantly between 200 and 100~K
(Fig.5). On the basis of considerations above concerning $1/q(T)$
and $I_0(T)$ being governed by $\rho(T)$, one therefore concludes
that with decreasing temperature the density of electronic
excitations coupled to this phonon also decreases, concurrently to
the decrease of the electronic background $B(\omega,T)$ already
noticed. Such a temperature behavior of the electronic Raman
scattering is most consistent with an opening of a gap in the
electronic excitations as the pseudogap in the underdoped high
temperature superconductors.\cite{timusk1999,letacon2006,hinkov2007}
It is worth noting here that the characteristic temperature T$^*
\approx 150$~K, below which Opel at al.\cite{opel2000} have observed
weak spectral weight loss in the $z(XY)\bar{z}$ spectra of twinned
films of YBa$_2$Cu$_3$O$_{6.5}$ (T$c \approx 60$~K), is in the same
temperature range.

 \begin{figure}[htbp]
\includegraphics[width=6cm]{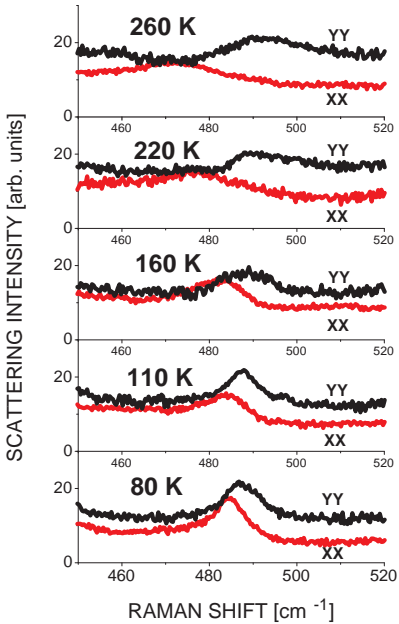}
\caption{(Color online) Variation with $T$ of the profile of the Raman
line associated with $A_g$ apex oxygen mode.}
\end{figure}

\section{Conclusions}
The experimental results reported here strongly suggest that some of
the previous Raman scattering
data\cite{misochko1994,palles1996,opel2000} obtained from aged
twinned surfaces of YBCO$_x$ ($x \approx 6.5$) samples may not be
representative of the ordered Ortho-II structure. This study
provides more reliable data for identification of the Ortho-II Raman
modes, as well as information on the variation of electronic scattering
and electron-phonon interactions in the temperature range where
opening of a pseudogap has been claimed.

\acknowledgments This work is supported in part by the State of
Texas through the Texas Center for Superconductivity at the
University of Houston (TcSUH) and by the U.S. Air Force Office of Scientific Research
(SPRING award ID FA9550-06-1-0401). The work at the University of Sherbrooke has been
supported by the National Science and Engineering Research Council of Canada, and the Fonds
Qu\'eb\'ecois de la Recherche sur la Nature et les Technologies.

\end{document}